\documentclass{emulateapj}

\newcommand{\ergss}{\ensuremath{{\rm erg\,s}^{-1}}}
\newcommand{\mc}{\multicolumn}
\newcommand{\rxjw}{RX~J1856.5$-$3754}
\newcommand{\rxjk}{RX~J0720.4$-$3125}
\newcommand{\RXJ}{RX~J0806.4$-$4123}
\newcommand{\rbs}{RX~J1308.6+2127}
\newcommand{\rxj}{RX~J0806}
\newcommand{\rbsb}{RX~J2143.0+0654}
\newcommand{\rxjvk}{RX~J1605.3+3249}

\newcommand{\xmm}{{\em XMM}}
\defcitealias{kvk05}{KvK05a}
\defcitealias{kvk05b}{KvK05b}
\defcitealias{vkk08}{vKK08}
\defcitealias{kvk09}{KvK09}
\defcitealias{hmz+04}{H+04}
\newcommand{\Hzsec}{\ensuremath{{\rm Hz}\,{\rm s}^{-1}}}
\newcommand{\secsec}{\ensuremath{{\rm s}\,{\rm s}^{-1}}}
\newcommand{\expnt}[2]{\ensuremath{#1 \times 10^{#2}}}   % scientific notation
\newcommand{\NH}{\ensuremath{N_{\rm H}}}

\begin{document}

\shorttitle{The Spin-down of the Nearby Isolated Neutron Star \RXJ}
\shortauthors{Kaplan \& van~Kerkwijk}

\title{Constraining the Spin-down of the Nearby Isolated Neutron Star
  \RXJ, and Implications for the Population of Nearby Neutron Stars}

\author{D.~L.~Kaplan\altaffilmark{1} and M.~H.~van  Kerkwijk\altaffilmark{2}}
\altaffiltext{1}{Hubble Fellow; KITP, Kohn Hall, University of
  California, Santa Barbara, CA 93106; dkaplan@kitp.ucsb.edu}

\altaffiltext{2}{Department of Astronomy and Astrophysics, University
  of Toronto, 50 St.\ George Street, Toronto, ON M5S 3H4, Canada;
  mhvk@astro.utoronto.ca}

\slugcomment{Accepted for publication in ApJ}
\begin{abstract}
  The nearby isolated neutron stars are a group of seven relatively
  slowly rotating neutron stars that show thermal X-ray spectra, most
  with broad absorption features.  They are interesting both because
  they may allow one to determine fundamental neutron-star properties
  by modeling their spectra, and because they appear to be a large
  fraction of the overall neutron-star population.  Here, we describe
  a series of \xmm\textit{-Newton}\ observations of the nearby
  isolated neutron star \RXJ, taken as part of larger program of
  timing studies.  From these, we limit the spin-down rate to $\dot
  \nu=\expnt{(-4.3\pm2.3)}{-16}\,\Hzsec$.  This constrains the dipole
  magnetic field to be $<\!\expnt{3.7}{13}\,$G at 2$\sigma$,
  significantly less than the field of $\sim\!10^{14}\,$G implied by
  simple models for the X-ray absorption found at 0.45\,keV.  We
  confirm that the spectrum is thermal and stable (to within a few
  percent), but find that the 0.45\,keV absorption feature is 
  broader and more complex than previously thought.  Considering the
  population of isolated neutron stars, we find that magnetic field
  decay from an initial field of $\lesssim\!\expnt{3}{14}\,$G accounts
  most naturally for their timing and spectral properties, both
  qualitatively and in the context of the models for field decay of
  Pons and collaborators.
\end{abstract}


\section{Introduction}
The so-called isolated neutron stars (INS; see \citealt{haberl07} and
\citealt{kaplan08} for reviews) are a group of seven (confirmed)
nearby ($\lesssim\!1\,$kpc) neutron stars with low
($\sim\!10^{32}{\rm\,erg\,s^{-1}}$) X-ray luminosities and long
(3--11\,s) spin periods.  They stand out from the normal neutron star
population because of their timing properties (which should not have
influenced their method of discovery via soft X-ray emission, although
see \citealt{hk98}).  The X-ray luminosities are consistent with
cooling neutron stars of ages $\sim 0.5\,$Myr \citep{plps04}, in rough
agreement with their kinematic ages (\citealt{wal01};
\citealt*{kvka02}; \citealt{msh+05,kvka07,mph+09}).

A crucial unknown in understanding the INS is their magnetic field
strengths.  This is important in understanding the X-ray spectra
(\citealt*{ztd04,mzh03}; \citealt{hkc+07,vkk07,haberl07}) and their
thermal history \citep{hk98}.  The X-ray spectra of the INS appear
thermal, with temperatures of $\sim\!10^6\,$K and with, in all but one
source, broad absorption features at energies of 0.2 to 0.75\,keV.
These features may give clues to the chemical composition and structure of the
INS surfaces, but also depend strongly on the magnetic field.  For a
hydrogen atmosphere the simplest possibilities are electron and proton
cyclotron resonances or transitions between bound states of neutral
hydrogen, all of which depend on the field strength.  Even with
different compositions (or states) the field strength is still
relevant.  At the same time, the magnetic field can affect the cooling
rate and thermal content of the neutron star atmosphere \citep*[e.g.,][
and references therein]{pmg09}.

Dipolar magnetic field strengths can be estimated
from coherent timing solutions, and we used X-ray observations to
derive such solutions for four INS, finding magnetic fields of
$\expnt{(1-3)}{13}\,$G \citep[][ hereafter \citetalias{kvk05},b,
\citetalias{vkk08}, \citetalias{kvk09}; also see
\citealt{vkkpm07}]{kvk05,kvk05b,vkk08,kvk09}.
Here, we constrain the spin-down rate and hence magnetic field
strength of the INS \object[RX J0806.4-4123]{\RXJ}\ (hereafter \rxj)
with a series of dedicated \xmm-\textit{Newton}\ observations.  We
also present a preliminary spectral analysis, but defer a detailed
phase-resolved analysis to a later paper.  

\rxj\ was identified as a
possible neutron star by \citet*{hmp98} on the basis of a soft thermal
spectrum and the absence of an optical counterpart.  Using \xmm,
\citet{hz02} confirmed that the spectrum was soft and blackbody-like
and identified a candidate $11.37\,$s periodicity.  Further
observations confirmed this periodicity and also suggested that the
X-ray spectrum was not purely a blackbody, but had a broad absorption
feature at $\approx\!0.45\,$keV (\citealt{hmz+04}; hereafter
\citetalias{hmz+04}).  In what follows, we assume a distance of
250\,pc to \rxj, derived by \citet{pph+07} from a comparison of the
X-ray absorption column density with a
model for the interstellar medium (using
$\NH=1.0\times10^{20}{\rm\,cm^{-2}}$, consistent with what we infer). 

With \rxj, we now have spin-down measurements or constraints for five of
the seven confirmed INS.  Examining the global properties of the INS
relative to other related neutron star populations, we attempt to
understand how the INS fit with the other groups.

The structure of this paper is as follows.  In Section~\ref{sec:obs},
we present our new data and perform our timing (\S~\ref{sec:timing})
and spectroscopic (\S~\ref{sec:spectra}) analyses.  We then consider
the implications of these results.  In \S~\ref{sec:insspectra}, we compare the
timing and spectral properties of \rxj\ to the rest of the INS and try
to understand the origin of the X-ray absorption features.  In
\S~\ref{sec:pop}, we extend the comparison to the pulsar population as
a whole, exploring the qualitative relation between the INS and
relevant sub-populations of pulsars.  From this, we are led to consider
the role that magnetic field decay may play in the evolution of the
INS, and in \S~\ref{sec:decay} we discuss our results in the context of a
specific model for field decay, that of \citet{pmg09}.  Finally, we
conclude in \S~\ref{sec:conc}.

\begin{deluxetable}{c c c c c c}
\tablewidth{0pt}
\tablecaption{Log of Observations and Times of Arrival\label{tab:obs}}
\tablehead{
&&\colhead{Exp.\tablenotemark{a}}&&\colhead{$f_{\rm bg}$\tablenotemark{a}}  & \colhead{TOA\tablenotemark{b}}\\
\colhead{Rev.}&
\colhead{Date}&\colhead{(ks)}&\colhead{Counts\tablenotemark{a}}&(\%)&\colhead{(MJD)}\\[-2.2ex]
}
\startdata
\dataset[ADS/XMM\#0106260201]{\phn168} &2000~Nov~08	& 15.6   & 25,604 & 1.6 &51856.691599(3) \\
\dataset[ADS/XMM\#0141750501]{\phn618} &2003~Apr~24	& 22.0   & 37,300 & 1.5 &52753.747077(3) \\
\dataset[ADS/XMM\#0552210201]{1542} &	2008~May~11	& \phn9.0 &11,467 & 4.1 &54597.501552(4) \\
\dataset[ADS/XMM\#0552210301]{1544} &	2008~May~15	& 10.0    &12,590 & 4.4 &54601.309174(4) \\
\dataset[ADS/XMM\#0552210401]{1551} &	2008~May~29	& \phn5.0 &\phn7,107 & 4.8 & 54615.274494(6) \\
\dataset[ADS/XMM\#0552210501]{1562} &	2008~Jun~20	& 13.0    &18,016 & 9.2&54637.594919(5)\\
\dataset[ADS/XMM\#0552210601]{1621} &	2008~Oct~15	& \phn9.0 &12,224 & 4.2 &54754.488408(4) \\
\dataset[ADS/XMM\#0552210901]{1631} &	2008~Nov~04	& \phn5.0 &\phn7,272 & 4.7 & 54774.185892(8) \\
\dataset[ADS/XMM\#0552211501]{1633} &	2008~Nov~09	& \phn6.0 &\phn9,386 & 27.8 &54779.285979(6) \\
\dataset[ADS/XMM\#0552211001]{1649} &	2008~Dec~10	& \phn9.0 &12,427 & 4.4 &54810.460578(4) \\
\dataset[ADS/XMM\#0552211101]{1705} &	2009~Mar~31	& \phn9.0 &11,185 & 4.6 &54921.908923(4) \\
\dataset[ADS/XMM\#0552211201]{1705} &	2009~Apr~01	& 26.0    &41,755 & 23.7&54922.901462(4)\\
\dataset[ADS/XMM\#0552211601]{1710} &	2009~Apr~11	& \phn8.0 &11,027 & 5.8 &54932.057781(4) \\
\enddata
\tablecomments{All observations used  the small window
  mode and  thin filter for  both EPIC-pn and EPIC-MOS1/2,
  except for Revs.~168 and 618, in which the full window mode was
  used, which meant that only the EPIC-pn data were suitable for timing.}
\tablenotetext{a}{The exposure time, number of counts, and estimated
  fraction of events due to background $f_{\rm bg}$ given here
  are for EPIC-pn only.}
\tablenotetext{b}{The TOA is defined as the time of maximum light of
  the fundamental (following Eqn.~\ref{eqn:pulse}) closest to the middle of each observation computed
  from the combined EPIC-pn and EPIC-MOS1/2 datasets, and is given
  with 1-$\sigma$ uncertainties. }
\end{deluxetable}

\section{Observations \& Analysis}
\label{sec:obs}
We observed \rxj\ eleven times with \xmm\ \citep{jla+01} in 2008 and
2009, and focus here on the data taken with the European Photon
Imaging Camera (EPIC) with pn and MOS detectors, all used in small
window mode with thin filters (Table~\ref{tab:obs}).  All our
observations, as well as those from \citet{hz02} and
\citetalias{hmz+04} (taken with the same filter, but with the full
window mode instead), were processed with SAS version 8.0.1.  We used
{\tt epchain} and {\tt emchain} and selected source events from a
circular region of $37\farcs5$ radius with energies between 150\,eV
(the default) and 1.2\,keV (where flares are minimized; the source is
not detected above 1.2\,keV, and the background grows increasingly
dominant below 150\,eV).  As recommended, we included only one and
two-pixel (single and double patterns 0--4) events with no warning
flags for pn, and single, double, and triple events (patterns 0--12)
with the default flag mask for MOS1/2.
We barycentered the event times using the {\em Chandra X-ray
Observatory} position from \citetalias{hmz+04}: $\alpha=08^{\rm
h}06^{\rm m}23\fs40$ and $\delta=-41\degr22\arcmin30\farcs9$ (J2000).
We extracted background lightcurves for pn from similarly sized
regions offset from the source, but at the same
\texttt{RAWY} coordinate, as recommended by the SAS User
Guide.\footnote{See
\url{http://xmm.esac.esa.int/external/xmm\_user\_support/documentation/sas\_usg/USG/node64.html}.}
For MOS1/2, the small-window mode does not permit such large background
areas, but we used several smaller areas to compensate.  The
background rate for our data (up to $\sim\!24$\% in one case) was
higher than for the archival data ($\sim\!2$\%), in some cases due
to severe flares.  However, restricting the time ranges to remove the
flares did not significantly alter our results or precision.

\begin{figure}
% timing/plot_total_profile.m
%\plotone{total_profile.eps}
\plotone{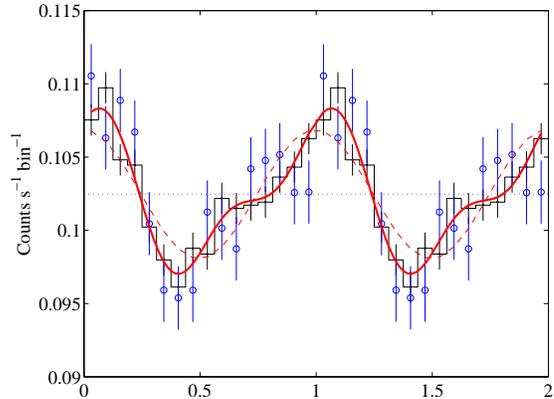}
\caption{Pulse profile of \rxj, repeated twice for clarity.  We show
the background-subtracted EPIC count-rate in each of 16 bins as a
function of pulse phase.  The solid curve is the best-fit model
including the first harmonic, while for comparison we also draw the
best-fit sinusoid as the dashed curve.  The histogram/points include
all events from our 2008--2009 observations phased up following
Table~\ref{tab:ephem}, while the open circles are the longest single
observation (Rev.~618 from 2003).}
\label{fig:pulse}
\end{figure}

\subsection{Timing Analysis}
\label{sec:timing}
As a starting place, we first determined the frequency that maximized
the power in a $Z_1^2$ periodogram for the EPIC-pn data from Rev.~1562
(the observation in 2008--2009 with the highest $Z_1^2$ power).  We
found a best-fit frequency of $\nu=0.087958\pm0.000006\,$Hz,
consistent with that found by \citetalias{hmz+04} for the earlier
data. A $Z_2^2$ periodogram that incorporates the first harmonic (see
below) gives the same result with slightly better precision, where the
uncertainty on $\nu$ is found with the same analytic expression as for
the $Z_1^2$ result (based on \citealt*{rem02}) and we have verified
with simulations that this is correct.

Using the above frequency, we determined the times-of-arrival (TOAs;
see Table~\ref{tab:obs}) for the combined EPIC data from each
observation by fitting the binned lightcurves
(following \citetalias{kvk05b}; we verified that we obtained similar results from unbinned
fits following \citealt{cash79}).  Like \citet{hz02}, we found
that the lightcurve of \rxj\
was best described not by a single sinusoid but instead by a sinusoid
and its first harmonic, where the number of counts in bin $i$ are:
\begin{equation}
N_i = A\left\{\cos\left[2\pi(f_0 t_i -\phi_0)\right] +
r_2\cos\left[4\pi(f_0 t_i-\phi_0-\Delta \phi_2)\right]\right\} + C
\label{eqn:pulse}
\end{equation}
where $f_0$ is the frequency of the fundamental from above, $t_i$ is
the time of bin $i$ on the interval $[0,1/f_0)$, and the fit parameters
are amplitude $A$, phase $\phi_0$, amplitude ratio $r_2$, phase offset
$\Delta \phi_2$, and constant rate $C$, similar to
\citetalias{kvk05b}.  We found initial best-fit parameters from 
the longer observations, finding that all
were reasonably consistent with a constant pulse profile
with $A/C\approx 4.6$\%, $r_2\approx 0.45$, and $\Delta \phi_2\approx
0.11$ (corrected for background).  This is consistent with the
$\sim\!6$\% pulsed fraction found by \citetalias{hmz+04}.  We refined
the parameters using the combined event list
derived from all 2008--2009 observations phased together (see below).

The spacing and precision of the TOAs is insufficient for an
unambiguous timing solution (unlike in \citetalias{kvk05},b but like
in \citetalias{vkk08} and \citetalias{kvk09}).  Instead we searched
for possible coherent timing solutions by iteratively trying sets of
cycle counts between the different TOAs from 2008 and 2009 (similar to
\citetalias{vkk08}, although as with \citetalias{kvk09} we did not
incorporate frequency information from each observation because it
does not add extra information).  We limit solutions to
$|\dot \nu|\lesssim\expnt{1.5}{-13}\,\Hzsec$, the 3-$\sigma$
incoherent limit from \citetalias{hmz+04}.

\begin{figure}
% timing/plot_soln.m
%\plotone{0806_residuals.eps}
\plotone{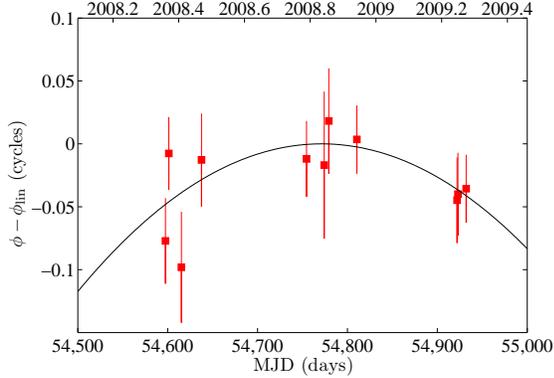}
\caption{Phase residuals for \rxj.  We show the residuals
  relative to a linear model ($\dot\nu=0$).  The line shows
  the best-fit quadratic solution.}
\label{fig:resids}
\end{figure}

We find one solution that is considerably
better than the alternatives with $\chi^2=5.1$ for 8 degrees of
freedom (DOF) and a small, not-quite significant spin-down rate of
$\dot \nu=\expnt{(-4.3\pm2.3)}{-16}\,\Hzsec$ (see Fig.~\ref{fig:resids},
Table~\ref{tab:ephem}).  This gives a 2\,$\sigma$ limit on the
magnetic field of field of $B_{\rm dip}<\expnt{3.7}{13}\,$G.  The next
best solution has $\chi^2=27.2$ and a very different spin-down rate:
$\expnt{(+4.32\pm0.02)}{-14}\,\Hzsec$. For a fit with 3 free
parameters, a change in $\chi^2$ of 22.1 means that the best-fit
solution is favored at 99.994\% confidence.  Even less likely
solutions are found for combinations of other cycle counts.

We confirmed this solution using a coherent $Z_1^2$ periodogram as a
function of both $\nu$ and $\dot \nu$.  The best solution was
consistent with that found in the TOA analysis, although it varied by
$\sim\!1\,\sigma$ in  $\dot \nu$.  This is likely because
the $Z_1^2$ did not incorporate any information about the harmonic,
while the TOA analysis did.  The best-fit peak has $Z_1^2=162.3$,
consistent with the background-corrected rms pulsed fraction of
$\approx\!3.8$\%.  Spin-down is not detected with  $Z_1^2(\dot
\nu=0)=162.2$, as the $Z_1^2$ does not include the harmonic.  Including
the harmonic at arbitrary phase and amplitude we get $Z_2^2=206.9$ at
$(\nu,\dot\nu)=(0.0879477628\pm\expnt{8}{-10}\,{\rm
Hz},\expnt{(-2.0\pm3.6)}{-16}\,\Hzsec)$, very close to the location of
the coherent solution (the increased uncertainty comes from not
including constraints on the pulse profile).

\begin{deluxetable}{cc}
\tablewidth{0pt}
\tablecaption{Measured and Derived Timing Parameters for \RXJ\label{tab:ephem}}
\tablehead{
\colhead{Quantity} & Value \\
}
\startdata
Dates (MJD) \dotfill & 54,598--54,932\\
$t_{0}$ (MJD)\dotfill        &54771.319605(2)\\
$\nu$ (Hz) \dotfill          &0.0879477624(9)\\
$\dot \nu$ ($10^{-16}\,$\Hzsec)&$-4.3(23)$\\
TOA rms (s) \dotfill         & 0.4 \\
$\chi^2$/DOF \dotfill        & 5.1/8\\
$P$ (s)\dotfill              & 11.37038592(12)\\
$\dot P$ ($10^{-14}\,$\secsec)\dotfill   & 5.5(30)\\
$\tau_{\rm char}$ (Myr)\dotfill& 3.3\\
$B_{\rm dip}$ ($10^{13}\,$G) \dotfill   &2.5 \\
$\dot E$ ($10^{30}\,{\rm erg}\,{\rm s}^{-1}$)\dotfill&1.5\\
\enddata
\tablecomments{Quantities in parentheses are the formal 1-$\sigma$
  uncertainties on the last digit.  $\tau_{\rm char}=P/2{\dot P}$ is
  the characteristic age, assuming an initial spin period $P_0\ll P$
  and a constant magnetic field; $B_{\rm
    dip}=\expnt{3.2}{19}\sqrt{P{\dot P}}{\rm\,G}$ is the magnetic field
  inferred assuming spin-down by dipole radiation; $\dot
  E=\expnt{3.9}{46}\nu\dot\nu\,{\rm erg\,s^{-1}}$ is the spin-down luminosity.  In
  addition to the nominal values above, the 2-$\sigma$ limits to those
quantities (based on $\dot \nu>\expnt{-8.9}{-16}\,\Hzsec$) are $\tau_{\rm char}>1.5\,$Myr, $B_{\rm
  dip}<\expnt{3.7}{13}\,$G, and $\dot E<\expnt{2.8}{30}\,\ergss$.
 }
\end{deluxetable}

After performing our coherent timing solution, we made a pulse profile
using the events from all 2008--2009 observations
(Fig.~\ref{fig:pulse}).  We found that the shape was consistent with
what we had assumed above, but that changing the pulse shape
parameters within the measured uncertainties changed the timing
solution slightly, typically by 10\% in $\dot \nu$ (20\% of the
uncertainty on $\dot \nu$).  Similarly, changing the timing solution
within the uncertainties changed the fitted parameters slightly.
However, since all such changes were significantly less than the
uncertainties, we decided to choose an average set of pulse profile
parameters and use them throughout.  The final parameters that we used
were $A/C=0.046\pm0.004$, $r_2=0.45\pm0.10$, and $\Delta
\phi_2=0.11\pm0.02$ (i.e., our arrival times are based on fits in
which only $\phi_0$ and $C$ are free parameters).  These
values are consistent with the timing solution given above and in
Table~\ref{tab:ephem}.  The profile with the harmonic is preferred
greatly to just a sinusoid, with $\chi^2=43.8$ for 13 DOF (fundamental
only) vs.\ $\chi^2=14.2$ for 11 DOF (with the harmonic).  Changing
parameters slightly gives similar but slightly different results.  For
example, with $r_2=0.55$ instead of $0.45$, we find $\dot \nu$
decreases by 2\% which is 5\% of the uncertainty on $\dot \nu$.
Decreasing $A/C$ to 4\% increases $\dot \nu$ by 7\% (15\% of the
uncertainty on $\dot \nu$).  The energy range that we used also
affects our results slightly, but the solution remains consistent
overall.  For instance, we see slight differences if we restrict our
analysis to the energy range of 300--500\,eV, where the pulsations
from \rxj\ are the very strong and the spectrum shows X-ray absorption
(see below).  While the pulsation amplitude increases to $A/C=7$\%,
the timing solution is consistent with $\dot
\nu=\expnt{(-2.1\pm2.2)}{-16}\,\Hzsec$.

Unfortunately, we  cannot unambiguously extrapolate our
solution back to the 2000 and 2003 observations and thus infer a
precise spin-down rate: the gaps are so large that even at $1\,\sigma$,
the uncertainty on the cycle count from the $\dot \nu$ uncertainty is
$\onehalf \sigma_{\dot \nu} \Delta t^2=7.2$\,cycles (where $\Delta
t=\expnt{1.8}{8}$\,s is the gap between the reference time and Rev.~618).

\subsection{Spectroscopic Analysis}
\label{sec:spectra}
With $\sim\!3$ times longer total exposure time compared to
\citetalias{hmz+04}, we wished to see whether the basic spectral fits
of \citetalias{hmz+04} are still valid and to look for possible
long-term variability such as that found for \rxjk\ by
\citet{dvvmv04}.

\begin{deluxetable*}{l c c c c c c c c c r}
\tablewidth{0pt}
\tabletypesize{\footnotesize}
\tablecaption{Results of Spectroscopic Fits to the Combined 2008--2009 EPIC-pn Data\label{tab:spec}}
\tablehead{
\colhead{Model\tablenotemark{a}} & \colhead{$kT^{\infty\,{\rm b}}$} & \colhead{$R_{\rm BB}^{\infty}$\tablenotemark{b}} & \colhead{$N_{\rm H}$} &
\colhead{$E_1$\tablenotemark{c}} & \colhead{$A_1$\tablenotemark{c}} &
\colhead{$E_2$\tablenotemark{c}} & \colhead{$A_2$\tablenotemark{c}} 
& \colhead{$F_X$\tablenotemark{d}} & \colhead{$F_X^U$\tablenotemark{d}} & \colhead{$\chi^2$/DOF}\\
 & \colhead{(eV)} & \colhead{(km)} & \colhead{($10^{20}\,{\rm
    cm}^{-2}$)} & \colhead{(eV)} & & \colhead{(eV)}  & &
\mc{2}{c}{($10^{-12}\,{\rm ergs\,s}^{-1}\,{\rm cm}^{-2}$)} &
}
\startdata
%Blackbody &96.5(2) & 1.380(6) & 0.0 & \nodata & \nodata &
%\nodata & \nodata & 2.3 & 2.3 & 1039/473\\
% & 93.5(2) & 1.54(1) & 0.4 & \nodata & \nodata &
%\nodata & \nodata &   2.3 & 2.5 &1150/473\\
% & 89.6(2) & 1.81(1) & 1.0 & \nodata & \nodata &
%\nodata & \nodata & 2.2 & 2.8 &1748/473\\
%Blackbody-$\lambda$\tablenotemark{e} &93.5(2) & 1.55(1) & 0.4 & \nodata & \nodata &
%\nodata & \nodata &   2.3 & 2.5 &1142/473\\
% & 89.4(2) & 1.82(1) & 1.0 & \nodata & \nodata &
%\nodata & \nodata & 2.2 & 2.8 &1714/473\\
%1G & 90.3(2) & 1.89(1) & 1.0 & 489(3) & 0.266(7)\phn & \nodata &
%\nodata & 2.3 & 2.9&750/471\\
% EW1=56.6 eV
%1G & 91.3(5) & 1.81(5) & 0.85(8) & 494(4) & 0.243(14) & \nodata &
%\nodata &  2.3 & 2.8 &746/470\\
% EW1=51.8 eV
%1G-$\lambda$\tablenotemark{e} & 91.0(5) & 1.89(5) & 1.04(8) & 476(4) & 0.280(14) & \nodata &
%\nodata &  2.3 & 2.8 &666.5/470\\
%2G & 88.9(8) & 2.25(10) & 1.6(1)\phn & 461(4) & 0.411(20) & 689(8)\phn & 0.269(19)
%& 2.3 & 3.4 &598/468\\
%2G-$\lambda$\tablenotemark{e} & 91.0(8) & 2.05(10) & 1.4(1)\phn &
%461(4) & 0.379(20) & 706(8)\phn & 0.192(19) & 2.3 & 3.4 &574/468\\
% EW1=87.5, EW2=57.2
%3G & 91.1(9) & 1.95 & 0.7(2)\phn & 454(5) & 0.364(21) & 681(11) & 0.231(20)  & 230 & 0.490(90) &
%2.5 & 2.7& 583/467\\
% EW1=77.4, EW2=49.2, EW3=26.1
Blackbody & 95.0(2) &1.432(8) & 0.0 & \nodata & \nodata & \nodata & \nodata &2.3 &2.3 & 649/392\\
1G & 95.1(11) & 1.93(7) & 0.95(11) & 486(5) & 0.25(2) & \nodata & \nodata & 2.3 &2.9 & 482/389\\
2G & 87.2(11) & 2.39(15) & 1.7(2) & 460(5) & 0.42(3) & 693(12) &0.26(3) & 2.3 & 3.5 & 402/387\\
\enddata
\tablenotetext{a}{A blackbody modified by interstellar absorption,
  plus 0, 1, or 2 Gaussian absorption lines following Eqn.~\ref{eqn:gauss}.}
\tablenotetext{b}{The effective temperature and radius of the best-fit
  blackbody as seen by a distant observer.  The radius is scaled to a
  distance of 250\,pc.}
\tablenotetext{c}{The central energy and absorption depth of Gaussian
  absorption lines with fixed ${\rm FWHM}=200\,$eV.  The equivalent
  width is $A (\pi/4\ln 2)^{1/2} {\rm FWHM}=212.9 A\,$eV.}
\tablenotetext{d}{Absorbed and unabsorbed fluxes in the 0.2--2.0\,keV band.}
%\tablenotetext{e}{These models were actually fit in wavelength space,
%  although we report the results here in energy space.}
\tablecomments{Only the 0.15--1.2\,keV range was fitted.  
  Quantities in parentheses are the formal 1-$\sigma$
  uncertainties on the last digit.  Quantities without uncertainties
  were held fixed for that particular fit.}
\end{deluxetable*}

To do so we examined all EPIC-pn spectra of \rxj. 
(A full spectral analysis, including the EPIC-MOS and RGS data
and a phase-resolved analysis, is in progress.)  We
used the same source and background extraction regions as for the
timing analysis, created appropriate
response files, and binned the spectral files such that the
bin width was at least 25\,eV (about one third of the spectral
resolution) and the number of source plus background counts was at least~25.

\begin{figure}
% timing/compare_pn.m
%\plotone{pnrawspectra.eps}
\plotone{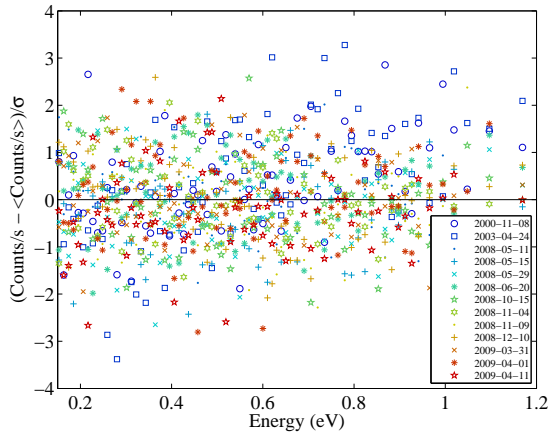}
\caption{Binned EPIC-pn spectra of \rxj, where we have subtracted off the
  mean spectrum and divided by the uncertainties.  The background has
  been subtracted, but no other corrections have been done.  The
  individual observations are labeled by their dates from Table~\ref{tab:obs}.
}
\label{fig:pnraw}
\end{figure}

We first compared the raw EPIC-pn spectra of all of the observations
against each other.  This did not include any response files or
calibration corrections, but even so the binned pn spectra were
generally consistent with each other implying no spectral change
(Fig.~\ref{fig:pnraw}).  The only small deviations were seen for the
archival data in 2000 and 2003 (reduced $\chi^2$ of 2--3), and much of
that could be corrected by a small slope across the 0.15--1.2\,keV
band: a fractional change of $\approx 0.15\,{\rm keV}^{-1}$ is
sufficient to get a reduced $\chi^2$ of 1--2 for all observations.
This change for the early data most likely reflects a small degree of event
pileup present in those observations above 0.7\,keV, caused by the 73\,ms
frame time of the full frame mode being insufficient to separate all
events (for the small window mode, with its 6\,ms frame time, pileup is not an issue).  Because of this
corruption, in the following fits we excluded the full frame data.
We note, however, that the results including those data were still
quite similar.

\begin{figure}
% timing/plot_pnspectra_abs.m
%\plotone{0806_spectum.eps}
\plotone{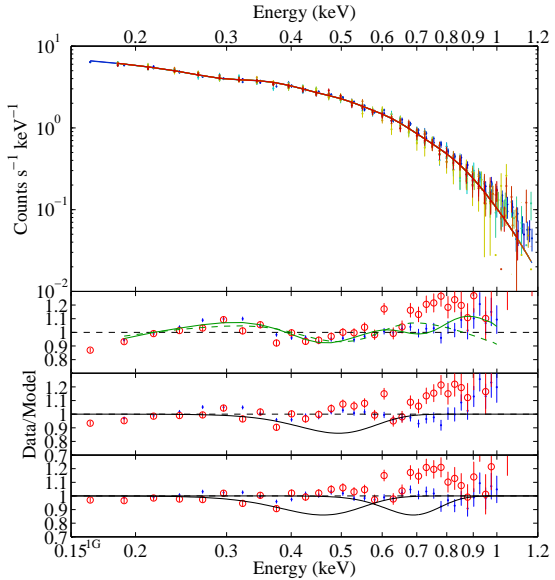}
\caption{EPIC-pn spectra of \rxj.  The top panel shows the data from
  each observation (as in Fig.~\ref{fig:pnraw}) along with the
  best-fit blackbody plus 2 absorption line model (2G; solid line).
  The lower panels give the ratio of the data from 2008--2009 (with
  all observations averaged together for clarify as the blue points,
  and Rev.~618 separately as the red circles) to the model for
  the best-fit blackbody, 1G, and 2G models (as labeled;
  Table~\ref{tab:spec}).  In the blackbody panel we plot the ratio of
  the models 
  1G/Blackbody (dashed green line) and 2G/Blackbody (solid green line).
  Where we have Gaussian absorption components, we also show the
  Gaussians schematically (black lines), although the amplitudes are
  on an arbitrary scale.  }
\label{fig:pnbb}
\end{figure}

We fit the pn data from 2008--2009 together using \texttt{sherpa}.  Like \citetalias{hmz+04},
we found that an absorbed blackbody did not provide a good fit.
Indeed, the deviations from a blackbody forced the interstellar
absorption column density \NH\ to zero.  Rather than setting \NH\
arbitrarily, we fit for an unabsorbed blackbody: while not realistic,
this provides a reference.  We list the corresponding radius and
temperature in Table~\ref{tab:spec}.  In Fig.~\ref{fig:pnbb}, one sees
that there are strong negative residuals near $0.45\,$keV.  We tried
fitting for this using Gaussian absorption lines, defined as follows,
\begin{equation}
F(E)=F_{C}(E)\left[1-\sum_{i=1}^{N_{\rm abs}} A_i
  \exp\left(-4 \ln 2 \frac{(E-E_i)^2}{{\rm FWHM_i}^2}\right)\right]
\label{eqn:gauss}
\end{equation}
where $F_C(E)$ is the continuum spectrum (a blackbody with
interstellar absorption), $E_i$, $A_i$, and FWHM$_i$ are the central
energy, absorption depth, and FWHM of absorption component $i$ and
the factor of $4\ln 2$ converts from a FWHM to a Gaussian $\sigma$.  

In Fig.~\ref{fig:pnbb}, one sees that the absorption is broad and
somewhat asymmetric.  We found that, as a result, if we left the FWHM
free, a single Gaussian became as wide as the blackbody peak.
Instead, therefore, we decided to fix the FWHM to 0.2\,keV, similar to
what was used by \citetalias{hmz+04} (0.16\,keV).  This is
arbitrary, but should at least give a qualitative sense of the amount
of absorption present.  Including this single component results in a
feature with an equivalent width of 53\,eV and a significantly
improved fit.  
As a consistency check, we kept the central energy and amplitude of
the absorption 
fixed and fit for the FWHM, finding a best-fit value of 0.23\,keV.

In Figure~\ref{fig:pnbb}, there still appear to be significant
deviations of the residuals of the 1G model near $0.65\,$keV, showing
that our single line could not reproduce the deviations from the black
body well.  We therefore added a second Gaussian absorption component
near that energy.\footnote{\citet{haberl07} discussed fitting two
lines in a fixed ratio, and that improved the quality of the fit
(albeit with less data).  However, it was not well constrained, and
both of lines from \citet{haberl07} would fit in the general region
where we find significant residuals.}  This significantly improved the
fit, with much flatter residuals and a reduced $\chi^2$ of 1.04.  The
change in $\chi^2$ is significant at the $\sim10^{-16}$ level,
according to an F-test.  We also fit Gaussians at a range of energies
across the 0.2--1.1\,keV band, and find that except for the edges of
the band (where the fitting breaks down) only lines near $0.68\,$keV
give significant improvements to the fit.  Compared to allowing the
FWHM of the first Gaussian to vary, we find that including a second
line gives a much better fit (e.g., reduced $\chi^2$ of 1.52 for
FWHM$_1=0.5\,$keV).

Even with two Gaussians, there is still structure in the residuals and
the reduced $\chi^2$ is not quite 1, likely reflecting the even-more
complex shape of the absorption feature.  The two lines in
Figure~\ref{fig:pnbb} overlap, and there are residuals at their edges.
The equivalent width of the first line has now increased to 89\,eV,
with the second at 55\,eV.  The energies of these lines are in an
almost 3:2 ratio which may reflect some underlying harmonic structure
to the absorption, but it may also just come from attempts to fit the
complex structure of the absorption with simple shapes: changing the
FWHM of each line changes the central energies and the ratio, which is
almost 1.7 for lines with FWHM of 0.3\,keV.  We could add additional
components to improve the fit but they become increasingly artificial.
We did try to fit in wavelength space where the shape of the
absorption is slightly different (e.g., \citealt{vkkd+04}).  There the
basic results hold, and while a single Gaussian does a somewhat better
job, it is still far from perfect. Overall the absorption seems truly
complex in shape, with a total equivalent width $\gtrsim140\,$eV.
Further phase-resolved fits may be able to help understand some of the
complexities, but we defer them to a later paper.

While the data were largely consistent with a constant spectrum, we
did one simple test in which we allowed the individual observations to
differ.  With the basic 2G fit from Table~\ref{tab:spec} the
observations had reduced $\chi^2$ ranging from 0.7 to 3.3, where the
longest observation (Rev.~618) had the worst $\chi^2$.  This is also
evident in Figure~\ref{fig:pnbb}, where the residuals from Rev.~618
are generally similar to the average residuals, with some slight
deviations at high energies.  With the exceptions of the data from
2000 and 2003, we find the data are consistent with a constant
spectrum.  The early data deviate slightly, giving slightly higher
temperature and lower radius, but again this is largely because of the
pileup-induced apparent hardening of the spectrum. Even including all
of the data in the fit, $kT$ changes by only a few eV with a resulting
change in the blackbody radius such that $R_{\rm BB}^2\propto kT_{\rm
BB}^{-4}$, i.e., the flux stayed relatively constant ($<2$\%).

\section{\RXJ\ and the INS}
\label{sec:discuss}
Much like the other INS for which we have derived timing
solutions/constraints, \rxj\ has a magnetic field near a few by
$10^{13}$\,G (nominally, $\expnt{2.5}{13}\,$G and
$<\!\expnt{3.7}{13}\,$G at 2\,$\sigma$) with a characteristic age of a
few Myr (nominally, $3.3\,$Myr and $>\!1.5\,$Myr at $2\sigma$) and a
spin-down luminosity of $\sim\!10^{30}\,\ergss$.  As such, the INS
form a very homogeneous group.  Here, we discuss the implications for
their spectral features, as well as for the overall neutron star
population.

\begin{figure}
% plot_ppdotnew.m
%\plotone{ppdot_all.eps}
\plotone{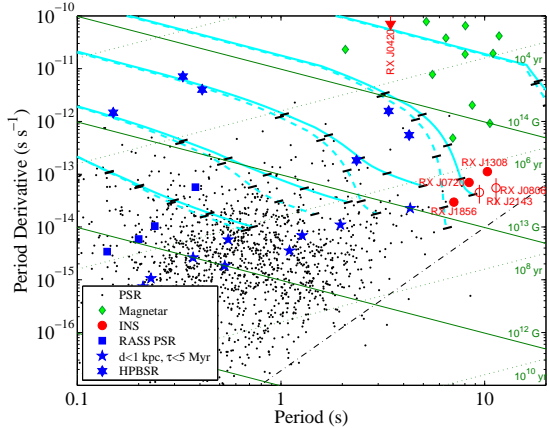}
\caption{$P$-$\dot P$ diagram.  We show the non-recycled pulsar
  population as points.  The individual types of objects listed in
  Table~\ref{tab:all} are also indicated: the INS are red circles (or
  upper limits where no $\dot P$ is known), the pulsars detected by
  \textit{ROSAT} are blue squares, the nearby/young pulsars are blue
  five-pointed stars, and the pulsars with dipole fields $\geq
  10^{13}$\,G and X-ray observations are blue six-pointed stars.  We
  also plot the magnetars \citep{wt06} as green diamonds. We indicate
  lines of constant dipolar field and characteristic age as labeled.
  Finally, we also show the results of evolutionary model from
  \citet{pmg09} as the thick lines (solid lines: dipole only; dashed
  lines: including a toroidal component with a strength of 50 times
  that of the dipole). The initial dipole fields are
  $10^{13,13.5,14,14.5,15}$\,G at the pole, there are cross-ticks at
  $(0.01,0.03,0.1,0.3,1)\,$Myr, we assumed initial periods of 0.1\,s,
  and we have divided the field at the pole by 2 to get the field at
  the equator for comparison with the spin-down field estimates
  \citep[see][]{lk04}.}
\label{fig:ppdot}
\end{figure}

\subsection{INS spectral features}
\label{sec:insspectra}

In \citet{vkk07} we compared the X-ray absorption of the different INS
to a simple model including ion cyclotron and neutral hydrogen
absorption in strong magnetic fields.  We noted that the absorption
energy of \rxj\ ($\approx\!0.45\,$keV) was similar to that of \rxjvk\
(\citealt{vkkd+04}), and we see here that the depths are also
comparable (we found 139\,eV for the equivalent width of the main
absorption line for \rxjvk).  \citet{haberl07} even found that a
second line with an energy 1.5 times the first significantly improved
the fit of \rxjvk, as we found here for \rxj, making the sources even
more similar, although we again caution that it is not clear how
unique or meaningful those fits are.  We speculated that the magnetic
fields might be the same, at $\sim\!10^{14}\,$G (this value is
required for both proton cyclotron and neutral hydrogen transitions;
also see \citealt{haberl07}).  However, even though we do not have a
clear detection of spin-down and thus cannot compare in detail, we can
exclude this picture for \rxj: our limits show that at least the
dipole component is quite a bit lower (see \citetalias{kvk09} for a
similar discussion regarding \rbsb).  The possible presence of
multiple lines complicates the situation somewhat, since the
``fundamental'' may be at lower energies and thus the inferred
magnetic field weaker.  However, since harmonics should be
significantly weaker for ion cyclotron resonance at least
\citep{pp76}, this seems unlikely to be the explanation.

Presuming that the absorption comes from some transition in the
atmosphere, there are a number of alternative explanations.  For one,
the magnetic field geometry could play a  role.  The dipole
component that we measure is only a projection of the true dipole
field, and the field on the surface could have higher order or
substantial toroidal components \citep{braithwaite09}.  Beyond that,
different INS could have different chemical composition, although then
we have to understand how the apparent emission radii of the INS are
relatively similar \citep{kvka07}.

Including \rxj\ in the magnetic field-effective temperature plane, it
is consistent with the line defined by \rxjw\, \rxjk, and \rbs\
\citepalias{kvk09}.  As discussed in \citetalias{kvk09}, this
correlation, while quite possibly a coincidence, may have some
relation to the origin of the INS and the coupling of the magnetic and
thermal evolutions (see below).  It is also possible that it is
influenced by surface condensation\footnote{In our discussion of the
effects of condensation in \citetalias{kvk09}, we incorrectly stated
that a condensed surface would inhibit a vacuum gap and hence radio
emission: in fact it is the opposite (Z.~Medin 2009, priv.~comm.).}
\citep{ml07b}.

\subsection{The INS, and the Neutron Star Population}
\label{sec:pop}
While the INS were discovered over 10 years ago, we still lack a
detailed appreciation for their place in the overall neutron star
population and for what makes them unique (although see
\citealt{pct+00,pcp+00,ptp06,pph+08} and references therein for some discussion).  A
full understanding requires detailed models of the birth and evolution of neutron
stars, but given the small numbers of objects and the large number of
free parameters, getting reliable constraints is necessarily difficult.  However, we
can gain some insight by comparing the INS to different related
neutron star populations.  For this purpose, we give the salient properties of different
objects in Table~\ref{tab:all}.  Aside from the seven confirmed INS,
we consider:
\begin{itemize}
\item Rotation-powered pulsars detected in the \textit{ROSAT} All-sky
  Survey with count-rates $\geq\!0.05\,{\rm s}^{-1}$ in the PSPC (referred
  to here as ``RASSPSRs'').
\item Other relatively young and nearby rotation-powered pulsars
  (referred to here as ``NearPSRs''; we
  adopted a distance limit of 1\,kpc and a characteristic age limit of
  5\,Myr).
\item Rotation-powered pulsars with dipole magnetic field
  $\geq\!10^{13}\,$G (so called high-$B$ pulsars, or ``HBPSRs''),
  but limited to those with X-ray observations.
\end{itemize}
These objects each make appropriate comparisons with the INS.  The
RASSPSRs are generally young and nearby neutron stars (we exclude the
Crab pulsar 
from further consideration as it is much younger and more distant than
the rest of the RASSPSRs), with X-ray
luminosities and distances similar to those of the INS.  Moreover, these were all detected
in the same survey that discovered the INS.  However, we do not know
exactly how young the INS are.  We therefore also include the
NearPSRs: a sample of moderately young, moderately nearby pulsars
regardless of their X-ray luminosity.  While the INS are likely within
500 or 700\,pc, we extend our range to 1\,kpc for the pulsars since the
distances of both classes are uncertain.  Similarly, we extend to
characteristic ages of 5\,Myr, compared to characteristic ages of
$\sim\!3\,$Myr for the INS and kinematic ages of $<\!1\,$Myr.
Finally, as the INS have larger than average magnetic fields, we also
include those HBPSRs that have comparable dipole fields
($\geq\!10^{13}\,$G) and X-ray observations. 

For each object, we give:
\begin{itemize}
\item The timing properties, comprising the spin-period and the
  properties derived from that and $\dot P$: dipole magnetic field,
  characteristic age, and spin-down luminosity.  With the exception of
  \rbsb\ and \rxj\ (where the spin-down measurements are only
  marginally significant) these data are uniformly of high quality.
\item The X-ray spectral properties, comprising the count-rate (as
  detected in the \textit{ROSAT} PSPC), temperature and radius of the
  best-fit blackbody, luminosity of the blackbody, and total
  luminosity including any non-thermal components.  The quality of
  these parameters varies by object.  For the INS they are relatively
  uniform, but the details of the spectral fit depend on the
  observations and the object as different objects have different
  levels of X-ray absorption (for example see \citealt{vkk07},
  \S\S~\ref{sec:spectra} and \ref{sec:insspectra}).  For the
  RASSPSRs the thermal components are from just one possible
  decomposition, and often two blackbodies are required in addition to
  a non-thermal component.  For the fainter NearPSRs and HBPSRs again
  there are problems of decomposition (now limited by signal-to-noise
  ratio), and in many cases we only have upper limits.
\item The distance (measured by astrometry where available) and any
  other age indicator (kinematic age or age of associated supernova
  remnant).  Distances from astrometry are ideal, but in
  marginally-significant cases (such as \rxjk; \citealt{kvka07})
  Lutz-Kelker bias \citep{lk73,smith03} can make them appear closer
  than they are, although other information such as X-ray absorption
  and Galactic geometry can be incorporated to improve the situation.
  For objects without astrometry, we rely on the dispersion measure
  (in radio) and hydrogen column densities (in X-rays) along with
  models for the Galaxy.  In both cases, the results can be
  unreliable, especially for close objects (as these are), although it
  is less likely that they are systematically biased.  Furthermore,
  the X-ray absorption column density can be unreliable as it is often
  covariant with other fitting parameters as well as with the assumed
  shape of the spectrum (e.g., \citealt{dvk06}).
\end{itemize}
For additional notes on the data, see Table~\ref{tab:all}.  

If we take the RASSPSRs and INS as a single sample detected by
\textit{ROSAT}, the two groups have roughly comparable sizes
\citep{pcp+03,kaplan04,kaplan08}.  Based on their temperatures and
independent ages at least some of the RASSPSRs may be slightly younger
than the INS, but the temperatures can be affected by non-thermal
emission processes absent in the INS.  Comparing the INS to the
RASSPSRs, the INS have comparable X-ray luminosities of
$\sim\!10^{32}\,\ergss$ \citep[e.g.,][]{pcp+03,kaplan04,kaplan08}.  The 
kinematic ages of the INS (when available) are comparable to if not
slightly larger than the ages of the RASSPSRs.  However, the
characteristic ages of the RASSPSRs are also small, $<\!1\,$Myr, while
the characteristic ages of the INS are all $>\!2\,$Myr.  All of the
RASSPSRs have periods $<\!0.4\,$s, while the INS have periods an order
of magnitude longer and considerably stronger magnetic
fields.\footnote{At some level, our comparison of magnetic fields,
pulse periods, characteristic ages, and spin-down luminosities is
degenerate since these parameters are all related.  Nonetheless, we
will continue to emphasize differences among multiple parameters when
appropriate.}  In contrast, for the slightly older NearPSRs the
characteristic ages are often comparable to those of the INS but the
X-ray luminosities are several orders of magnitude lower
($\sim\!10^{29}\,\ergss$).  The INS also have much lower $\dot E$ than any 
other population.  Finally, we note that INS have larger radii and smaller
effective temperatures for the same X-ray luminosity.  Is this an
innate difference, perhaps as a result of the magnetic field geometry?
Or is it artificial, perhaps a consequence of unrealistic emission models?

As we discussed previously \citepalias[e.g.,][]{kvk05}, we need to
reconcile the discrepant kinematic and characteristic ages of the INS.
For one or two objects it could be a coincidence, but even for the
objects without kinematic ages (\rxj, \rbsb), the luminosities suggest
the true ages are substantially shorter than the characteristic ages.
While having a very long initial period is possible, it would have to
be very close to the current value.  For other sources where the
characteristic ages exceed the true ages, like PSR~J0538+2817
\citep{klh+03,nrb+07}, the initial spin period required in order to
have spun-down to their current periods in their true ages (assuming
dipole braking) is $<\!0.2\,$.  This is longer than the
traditionally-assumed $\sim\!30\,$ms (\citealt*{lps93};
\citealt{mgb+02}), but similar to the more recently favored periods of
a few hundred ms (\citealt{klh+03,vml+04}; \citealt*{ghs05};
\citealt{fgk06}), and, most relevant for the present purposes, much
smaller than what would be required for the INS ($\gtrsim\!5\,$s).

\citet{klob+06} have discovered a pulsar whose
spin-down torque varies quasi-periodically along with its radio
emission.  For approximately 10--20\% of the time PSR~B1931+24 (an
otherwise ordinary pulsar with characteristic age of 1.6\,Myr and
dipole field of $\expnt{2.6}{12}$\,G) is visible as a radio pulsar,
but for the remaining time the radio emission is not detectable and
the torque is reduced to $\twothirds$ of its normal value.  The
difference is attributed to the presence (or strength) of an energetic
plasma wind, where a strong wind leads to radio emission and higher
torque.  While we suspect from its H$\alpha$ nebula that \rxjw\ does
indeed have an energetic wind and brakes by magnetic dipole radiation,
that scenario is not completely self-consistent (see below).  So the
absence of radio emission from the INS may point to cessation (or
at least diminution) of a wind and its associated torque, and this
could be a recent development in the histories of the sources if radio
emission has just shut off.  If the wind/radio emission only stopped
recently (the worst case scenario) then the period evolution would
have been dominated by a spin-down approximately $\slantfrac{3}{2}$
times what we see now and the characteristic age that we measure now
would be high by 50\%.  This would go some but not all of the way
toward resolving the discrepancy between the timing and kinematic
ages, although differences in geometry (also see \citetalias{kvk09})
could lead to slightly larger factors (other intermittent pulsars have
since been discovered with slightly larger ratios of 1.7 instead of
1.5, possibly due to differences in alignment between the rotation and
magnetic axes; \citealt{kramer08}).  Such a comparison also suggests
that there could be torque and/or radio flux variations in the INS,
neither of which has been seen (except for the one timing/spectral
change for \rxjk; \citealt{vkkpm07}) but the sampling has been very
sparse.

Another possibility is that spin-down does not follow the expectations
for magnetic dipole radiation.  For instance, a decaying magnetic
field would give rise to a situation like what we see for the INS, as
it would allow the sources to spin down rapidly early in their
evolution \citepalias{kvk09}.  Field decay was proposed previously for
the INS as a way to keep them hotter longer and thereby make them
overrepresented in a local sample \citep{hk98}.  Initial timing
results suggested that field decay was not presently heating the INS
\citep{zhc+02,kkvkm02}, and indeed the INS have roughly the expected
thermal luminosities for their kinematic ages (especially given the
large uncertainties and the steep decline in thermal luminosity for
objects of this age; see e.g., \citealt{plps04}).  However, as we
discuss below, field decay may have profoundly affected the magnetic
and rotational evolution of the INS, and may have some other visible
consequences.  The models of field decay used by \citet{kkvkm02} were
rudimentary, and including decay by other modes as well as a strong
toroidal component can greatly change the outcome.

Regardless of the specific model, if the characteristic ages are
correct for the INS, they are too luminous by more than a factor of
$100$ to be powered by residual heat.  They would therefore require an
extra energy source which we know cannot be the spin-down luminosity:
the remaining alternative is magnetic field decay.  If instead (and
as we believe) the kinematic ages are correct, then the X-ray
luminosities are reasonable, but we need to explain the long periods
(and the associated large characteristic ages, low spin-down
luminosities).  Again, magnetic field decay seems to be the best
option.

\subsection{The INS and a Model of Magnetic Field Decay}
\label{sec:decay}

The coupled magnetic and thermal evolution for neutron stars was
studied in detail by \citet{pmg09}.  These authors assumed a range of
initial field and temperature configurations and then followed them
over time.  Briefly, they found that for all sources with initial
magnetic fields $\gtrsim\!\expnt{5}{13}\,$G, the final magnetic field
is $\sim\!\expnt{3}{13}\,$G at 0.5\,Myr.  But for all sources with
weaker initial magnetic field the final magnetic field is just half
the initial field.  We note that the model of \citet{pmg09} assumes
magnetic field configurations and mechanisms of decay that may well be
overly simplified.  One of the biggest free parameters in their models
is the ratio of toroidal to poloidal fields.  This can vary
significantly \citep[e.g.,][]{braithwaite09}, and the non-linearity of
field decay means that a strong, decaying toroidal component can alter
the weaker poloidal component while being otherwise invisible.  

Despite the above uncertainties, it seems encouraging that the main
findings of \citet{pmg09} for the results of field decay are similar
to what we infer for the INS: neutron stars grouped around
$\expnt{2}{13}\,$G at true ages of $\sim\!0.5\,$Myr or older.  Taking the magnetic
field evolution and using it to infer the spin-down history, one
expects that neutron stars with strong fields quickly move across the
$P-\dot P$ diagram, losing memory of the initial spin period.  We find
that as long as the initial field is between $\expnt{1}{14}\,$G and
$\expnt{7}{14}\,$G, then the neutron star ends up at a period of 3 to
15\,s (Fig.~\ref{fig:ppdot}).  Coupling the thermal evolution in with
the magnetic field, \citet{pmg09} find luminosities of
$\sim\!10^{32}\,\ergss$ at ages of $\sim\!0.5\,$Myr, similar to what
is expected for less strongly magnetized neutron stars; the excess
energy from the field decay is mostly radiated at earlier ages, with
the thermal luminosity being substantially higher around $10^5$\,yr.
At $\sim\!0.5\,$Myr, the difference may still be a factor of a few
(dependent on field configuration and strength), but this would be
largely lost in the observational uncertainties (age, distance, etc.).
This small remaining difference should not lead to a great
over-representation in a local sample, and therefore the fact that
about half of the young neutron stars detected by \textit{ROSAT}\ (the
INS plus the RASSPSRs) have long periods (and presumably strong
fields) should give a reasonable clue to the true population.  In a
bit more detail, this will depend on the slope of the cooling curve:
for power-law cooling $L\propto t^{-\alpha}$, increasing the
luminosity by a factor of $\lambda$ will lead to an increase in
population size of $\lambda^{-1/\alpha}$ in a flux-limited sample if
the population is in a constant volume (as might be expected for soft
X-ray sources, since the exponential cutoff of interstellar absorption
limits their detectability to $\lesssim\!1\,$kpc).  If the volume can
increase too, we get an additional factor of $\lambda$ in the
population (assuming the population is confined to the Galactic disk).
The true situation will probably be in between these extremes.  With
$\alpha$ typically between 2 and 3 at a few $10^5$\,yr (for photon
cooling), a factor of at most a few increase in luminosity (due to
field decay) would lead to a factor of $\lesssim\!2$ increase in
population.  Therefore, the INS would represent between one quarter
and half the total population.

As an aside, we note that an interesting implication of the model of
\citet{pmg09} is that neutron stars may be hotter on the equator than
at the poles, a combination of Joule heating from field decay and
reduced conductivity preventing the heat from going inward.  While the
effect may be quite small at the ages of the INS, this may still
complicate the interpretation of the lightcurve and phase-resolved
spectroscopy of the INS.  Typically, a model with a warm pole and cool
equator is assumed \citep[e.g.,][]{br02,ho07}, although more
complicated models have also been considered \citep{zt06}.  If the
geometry were reversed that could lead to different interpretations,
and work is on-going to see if there are clear consequences to that
(W.~C.~G.~Ho, 2009, pers.\ comm.).  The asymmetry between equator and
pole is time- and field-dependent, being most apparent for strong
fields and at early times.  For weak fields, the standard hot poles
are regained.  It may be that the very low pulsed fractions and limits
found in sources like \rxjw\ and \rxjvk\ are a consequence of being
near the time where the surface becomes nearly isothermal, and that
this underlies the difficultly in finding geometries that satisfy the
observed pulsation limits \citep{br02,ho07}.  If there is still a
strong, buried toroidal field it could also lead to some ongoing,
low-level decay, which might give rise to some of the spectral and
temporal evolution seen in \rxjk\
\citep{dvvmv04,vdvmv04,htdv+06,vkkpm07}, although nothing similar has
been seen in the long-term monitoring of other sources
(\S~\ref{sec:insspectra};
\citealt{haberl07}; \citetalias{kvk09}).

If, as we posit, magnetic field decay has influenced the period
evolution of the INS, we can ask what the progenitors and descendants
of the INS might be.  \citet{hk98} discussed \rxjk\ as an old
magnetar.  This could still be the case, although based on the models
of \citet{pmg09} the initial magnetic fields for the INS would be
close to $\sim\!3\times10^{14}\,$G and not much higher and would have followed
an evolution that only skimmed the parameter range occupied by the
magnetars (see Fig.~\ref{fig:ppdot}).  The descendants of the stronger-field
magnetars would be expected to end up with longer periods.
The INS do not actually seem to be old
versions of at least the X-ray-bright HBPSRs.  These HBPSRs mostly
have $\dot E>10^{36}\,\ergss$ (although this may partly be a selection effect,
as this would increase $L_{\rm X}$).  Using the model of \citet{pmg09}
and evolving typical HBPSRs to ages of $0.5\,$Myr, we would expect
$\dot E\sim 10^{32}\,\ergss$, which exceeds by an order of magnitude
what the INS have.  These HBPSRs may in fact have magnetic fields
close to their initial values with little decay, while the INS started
with $\gtrsim\!10^{14}\,$G and have decayed.  A few of the HBPSRs,
like PSR~B0154+61 ($\log_{10}\dot E=32.8$, $\log_{10}L_{\rm X}<32.3$),
PSR~J1819$-$1458 ($\log_{10}\dot E=32.5$, $\log_{10}L_{\rm X}=33.7$),
and PSR~J1718$-$3718 ($\log_{10}\dot E=33.3$, $\log_{10}L_{\rm
X}\approx 33.5$), may be more similar, although we have no independent age
estimates for those objects.  Of these, the comparison to
PSR~J1819$-$1458 may be particularly interesting, as it emits only
sporadic radio bursts as a so-called ``Rotating RAdio Transient''
(RRAT; \citealt{mll+06}).  The possibilities of a connection between
the RRATs and the INS have already been discussed in several places
\citep[e.g.,][]{ptp06}, but the spectral similarities \citep{mrg+07}
and the recent detection of extended X-ray emission from
PSR~J1819$-$1458 that seems too large for its $\dot E$ \citep{rmg+09}
highlight it even further.  The connection can only go so far, though:
searches for RRAT-like radio emission from the INS have not been
successful \citep{kml+09}.

As for the descendants of the INS, while they might be too faint for
X-ray detection, they could still be apparent in radio surveys
(although there are observational selection effects against
long-period objects).  However, very few objects are known in that
part of the $P-\dot P$ diagram, and there are not enough to span the
expected ages of up to $10^7\,$yr.  Does this mean that the old INS
have very narrow radio beams that miss the Earth, therefore reducing
their prevalence in radio surveys?  Or are there no radio beams?  Will
the surfaces condense to allow the formation of vacuum gaps
(\citealt{ml07b}) or is this inhibited?  This is largely a function of
composition and state, which, as we discussed, remains uncertain.
The H$\alpha$ nebula around \rxjw\ might be taken as
evidence for the generation of energetic particles (\citealt{vkk01b,kvka02}), but the
required $\dot E$ is far higher than what is inferred from timing
\citep{vkk08}.

\defcitealias{kkvk02}{1}
\defcitealias{kvk05b}{2} 
\defcitealias{hsh+03}{3} 
\defcitealias{shhm05}{4} 
\defcitealias{shhm07}{5} 
\defcitealias{mph+09}{6} 
\defcitealias{kvkm+03}{7} 
\defcitealias{mzh03}{8} 
\defcitealias{kvk05}{9} 
\defcitealias{kvka07}{10} 
\defcitealias{vkkpm07}{11} 
\defcitealias{hztb04}{12} 
\defcitealias{hmz+04}{13} 
\defcitealias{haberl07}{14} 
\defcitealias{zct+01}{15} 
\defcitealias{zct+05}{16} 
\defcitealias{rtj+07}{17} 
\defcitealias{zmt+08}{18} 
\defcitealias{kvk09}{19} 
\defcitealias{bhn+03}{20} 
\defcitealias{vkk07}{21} 
\defcitealias{vkk08}{22} 
\defcitealias{kkvk03}{23} 
\defcitealias{vkkd+04}{24} 
\defcitealias{msh+05}{25}
\defcitealias{zdlmt06}{26} 
\defcitealias{wag+01}{27} 
\defcitealias{pzs+01}{28} 
\defcitealias{dlrm03}{29} 
\defcitealias{ms02}{30} 
\defcitealias{btgg03}{31} 
\defcitealias{dlcm+05}{32} 
\defcitealias{mgb+02}{33} 
\defcitealias{lll05}{34} 
\defcitealias{fwa07}{35} 
\defcitealias{kbm+03}{36} 
\defcitealias{rn03}{37} 
\defcitealias{nrb+07}{38} 
\defcitealias{crr+09}{39}
\defcitealias{cbv+09}{40} 
\defcitealias{bjk+05}{41}
\defcitealias{dtbr09}{42}
\defcitealias{ghm+08}{43} 
\defcitealias{ccv+04}{44} 
\defcitealias{mpg08}{45} 
\defcitealias{bt97}{46}
\defcitealias{bwt+04}{47} 
\defcitealias{zsp05}{48} 
\defcitealias{cgk+01}{49} 
\defcitealias{cmgc04}{50}
\defcitealias{gkc+05}{51} 
\defcitealias{shk08}{52} 
\defcitealias{gcm+95}{53} 
\defcitealias{gbm+99}{54} 
\defcitealias{gak+02}{55}
\defcitealias{lt08}{56}
\defcitealias{ggg+08}{57}
\defcitealias{ksh08}{58}
\defcitealias{gw03}{59}
\defcitealias{hsp+03}{60}
\defcitealias{lwa+02}{61}
\defcitealias{clb+02}{62}
\defcitealias{ltw08}{63}
\defcitealias{km05}{64}
\defcitealias{mrg+07}{65}
\defcitealias{gklp04}{66}

\setlength{\tabcolsep}{2pt}
\begin{deluxetable*}{lccccccccccccc}
\tablewidth{0pt}
\tabletypesize{\footnotesize}
\tablecaption{Properties of the Isolated Neutron Stars and Related
  Rotation-Powered Pulsars\label{tab:all}} 
\tablehead{
\colhead{Source} & \mc{4}{c}{Timing\tablenotemark{a}} && \mc{5}{c}{Spectrum\tablenotemark{b}}&
\colhead{d\tablenotemark{c}} & \colhead{Age\tablenotemark{d}} &
\colhead{Refs.}\\ \cline{2-5} \cline{7-11}
 & \colhead{$P$} & \colhead{$B_{\rm dip}$} &\colhead{$\tau_{\rm char}$} &
 \colhead{$\log_{10}\dot E$} && \colhead{PSPC} & \colhead{$kT$} & \colhead{$R_{\rm BB}$} &
 \colhead{$\log_{10} L_{\rm X,BB}$} & \colhead{$\log_{10} L_{\rm X,tot}$} \\
& \colhead{(s)} & \colhead{($10^{12}\,$G)} & \colhead{(Myr)} &
\colhead{($\ergss$)} &&
\colhead{(s$^{-1}$)}  & \colhead{(eV)} & \colhead{(km)} &
\colhead{(\ergss)} &\colhead{(\ergss)} & \colhead{(pc)} &  \colhead{(Myr)}\\
}
\setlength{\tabcolsep}{0pt}
\startdata
\mc{14}{l}{Isolated Neutron Stars}\\[0.5ex]
RX~J1308.6+2127 & 10.31 & 34 & 1.5 & 30.6 && 0.3 & 102 &4.1 &32.4 & 32.4
 & 500 & 0.8--1.2 & \citetalias{kkvk02,kvk05b,hsh+03,shhm05,shhm07,mph+09}\\
% \citenum{kkvk02,kvk05b,hsh+03,shhm05,shhm07,mph+09}
RX~J0720.4$-$3125 & 8.39 & 24 & 1.9 & 30.7 && 1.6 & 87 & 6.4 &32.5 & 32.5
 & 360 & 0.5--1.0 & \citetalias{kvkm+03,mzh03,kvk05,kvka07,vkkpm07,hztb04}\\
%\citenum{kvkm+03,mzh03,kvk05,kvka07,vkkpm07,hztb04}\\
RX~J0806.4$-$4123\tablenotemark{e} & 11.37 & 25 & 3.3 & 30.2 &&0.4 & 92 &1.3
 &31.2 & 31.2 & 250 & \nodata & \citetalias{hmz+04,haberl07}, this work\\
%\citenum{hmz+04,haberl07}
RX~J2143.0+0654 & 9.44 & 20 & 3.7 & 30.3 && 0.2 & 102 &3.2 & 32.1 & 32.1
 & 430 & \nodata & \citetalias{zct+01,zct+05,rtj+07,zmt+08,kvk09}\\
% \citenum{zct+01,zct+05,rtj+07,zmt+08,kvk09}
RX~J1856.5$-$3754 & 7.06 & 15 & 3.8 & 30.5 && 3.6 & 62 & 6.2 &31.9 & 31.9 &
 160 & 0.4 & \citetalias{bhn+03,vkk07,vkk08}\\
% \citenum{bhn+03,vkk07,vkk08}\\
RX~J1605.3+3249 & \nodata & \nodata & \nodata & \nodata&& 0.9 & 93 &4.7
 &32.3 & 32.3 & 390 & 0.1--1.0 & \citetalias{kkvk03,vkkd+04,haberl07,msh+05,zdlmt06}\\
% \citenum{kkvk03,vkkd+04,haberl07,msh+05,zdlmt06}\\
RX~J0420.0$-$5022 & 3.45 & \nodata & \nodata & \nodata &&0.1 & 45 &3.3 &
 30.8 & 30.8 & 345 & \nodata & \citetalias{hmz+04,haberl07}\\
% \citenum{hmz+04,haberl07}\\
\tableline
\mc{14}{l}{Rotation Powered Pulsars with ${\rm PSPC}>0.05\,{\rm
 s}^{-1}$ (RASSPSRs)}\\[0.5ex]

Crab & 0.03 & 3.8 & 0.0012 & 38.7 && 48.4 & \nodata & \nodata& \nodata & 37.6 &
 2000 & 0.001 & \citetalias{wag+01}\\
% \citenum{wag+01}\\
Vela & 0.09 & 3.4 & 0.011 & 36.8 && 3.4 & 128 & 2.1 & 32.3 & 32.8 & 287 & 0.01 &
 \citetalias{pzs+01,dlrm03}\\
PSR~B0656+14 & 0.38 & 4.7 & 0.11 & 34.6 && 1.92 & 56 & 20.9 & 32.7 & 32.7 & 288 &
 0.1 &  \citetalias{ms02,btgg03,dlcm+05}\\
PSR~B1951+32 & 0.04 & 4.9 & 0.11 & 36.6 && 0.07 & 130 & 2.1 & 32.2 & 33.3 & 2000
 & 0.06 & \citetalias{mgb+02,lll05}\\
Geminga & 0.24 & 1.6 & 0.3 & 34.5 && 0.54 & 43 & 8.6 & 31.9 & 31.8 &
250 & \nodata & \citetalias{dlcm+05,fwa07}\\
PSR~B1055$-$52 & 0.20 & 1.1 & 0.5 & 34.5 && 0.35 & 68 & 12.3 & 32.6 & 32.6 & 750
 & \nodata & \citetalias{kbm+03,dlcm+05}\\
PSR~J0538+2817 & 0.14 & 0.7 & 0.6 & 34.7 &&0.06 & 160 & 3.2 & 32.9 & 32.9
 & 1470 & 0.04 & \citetalias{rn03,nrb+07}\\
\tableline
\mc{14}{l}{Rotation Powered Pulsars with $d<1\,$kpc, $\tau_{\rm
 char}<5\,$Myr  (NearPSRs)}\\[0.5ex]

PSR~J1741$-$2054 & 0.41 & 2.7 & 0.4 & 34.0 && \nodata & \nodata & \nodata & \nodata & 31.1 & 400 & \nodata & \citetalias{crr+09}\\
PSR~B0450$-$18 & 0.55 & 1.8 & 1.5 & 33.1 &&  \nodata & \nodata &
\nodata & \nodata & \nodata & 760 & \nodata & \citetalias{cbv+09}\\
PSR~B0450+55 & 0.34 & 0.9 & 2.3 & 33.4 && \nodata & \nodata& \nodata& \nodata& \nodata& 1190 &
 \nodata & \citetalias{cbv+09}\\
PSR~J1918+1541 & 0.37 & 1.0 & 2.3 & 33.3 && \nodata &\nodata &\nodata &\nodata &\nodata & 680 &
 \nodata & \nodata \\
PSR~B0628$-$28 & 1.24 & 3.0 & 2.8 & 32.2 && 0.003 & \nodata & \nodata
& $<29.4$ & 30.1 & 332 & \nodata & \citetalias{bjk+05,dtbr09} \\
PSR~B2045-16 & 1.96 & 4.7 & 2.8 & 31.8 && \nodata &\nodata &\nodata &\nodata &\nodata & 950 &
 2 & \citetalias{cbv+09}\\
PSR~B1845$-$19 & 4.31 & 10.1 & 2.9 & 31.1 && \nodata &\nodata &\nodata &\nodata &\nodata & 950 &
 \nodata  & \nodata \\
PSR~B0834+06\tablenotemark{f} & 1.27 & 3.0 & 3.0 & 32.1 && \nodata &
 170 & 0.03 & 29.0 & 29.0 & 640 & \nodata & \citetalias{ghm+08}\\
PSR~B1929+10 & 0.23 & 0.5 & 3.1 & 33.6 && 0.012 & 300 & 0.03 & 30.0  & 30.4 & 361
 & 1--2 & \citetalias{ccv+04,mpg08}\\
PSR~J1908+0734 & 0.21 & 0.4 & 4.1 & 33.5 && $<0.005$ &\nodata &\nodata & \nodata &\nodata & 790 &
 \nodata &  \citetalias{bt97} \\
PSR~B0823+26 & 0.53 & 1.0 & 4.9 & 32.7 && 0.0016 & $<101$ &
 0.2\tablenotemark{g} & $<29.7$ & 29.4 & 360 & \nodata & \citetalias{bwt+04}\\
PSR~B0943+10\tablenotemark{h} & 1.10 & 2.0 & 5.0 & 32.0 && \nodata &
 270 & 0.02 & 28.7 & 28.7 & 630 & \nodata & \citetalias{zsp05}\\
\tableline
\mc{14}{l}{Pulsars With $B\geq 10^{13}\,$G and X-ray Observations (HBPSRs)}\\[0.5ex]
PSR~J1119$-$6127 & 0.41 & 41 & 0.0016 & 36.3 && \nodata & 210  &
 3.2 & 33.4 & 33.6 & 8400 & $<0.01$ &  \citetalias{cgk+01,cmgc04,gkc+05,shk08}\\
PSR~B1509$-$58 & 0.15 & 15 & 0.0016 & 37.3 && \nodata & \nodata &
 \nodata & \nodata & 34.3 & 5200 & 0.006--0.02 & \citetalias{gcm+95,gbm+99,gak+02}\\
PSR~J1846$-$0258\tablenotemark{i} & 0.33 & 49 & 0.0007 & 36.9 &&
\nodata & \nodata & \nodata & \nodata & 34.5 & 6000 & $<0.001$ &  \citetalias{lt08,ggg+08,ksh08}\\
PSR~J1124$-$5916 & 0.14 & 10 & 0.002 & 37.0 && \nodata & $<102$ & 12\tablenotemark{j}
 & $<33.3$ & 33.1 & 6000 & 0.03 & \citetalias{gw03,hsp+03}\\
PSR~J1930+1852 & 0.14 & 10 & 0.002 & 37.0 && \nodata & \nodata &
\nodata & \nodata & 34.6 & 6200 & \nodata & \citetalias{lwa+02,clb+02,ltw08}\\
PSR~J1718$-$3718 & 3.38 & 74 & 0.03 & 33.3 && \nodata & 145 & 7.7 &
 33.5 & 33.5 & 4500 & \nodata & \citetalias{km05}\\
PSR~J1819$-$1458 & 4.26 & 50 & 0.12 & 32.5 && \nodata & 140 & 11 & 33.7
& 33.7 & 3600 & \nodata & \citetalias{mrg+07}\\
PSR~B0154+61 & 2.35 & 21 & 0.2 & 32.8 && \nodata & $<63$ &
10\tablenotemark{j} & $<32.3$ & $<32.3$ & 1700 & \nodata & \citetalias{gklp04}\\
%PSR~J1814$-$1744 & 3.94 & 55 & 0.08 & 32.7 &&\\
%PSR~J1847$-$0130 & 6.71 & 94 & 0.08 & 32.3 &&\\
%\tableline
%PSR~J2144$-$3933  & 8.50 & 2.1 & 272 & 28.5 && \nodata & \nodata &
%\nodata & \nodata & \nodata & 180 & \nodata & \citenum{dsb+98,ymj99}\\
\enddata
\tablerefs{
\citetalias{kkvk02}:   \citet*{kkvk02};  
\citetalias{kvk05b}:   \citet{kvk05b};  
\citetalias{hsh+03}:   \citet{hsh+03};  
\citetalias{shhm05}:   \citet{shhm05};  
\citetalias{shhm07}:   \citet{shhm07};  
\citetalias{mph+09}:   \citet{mph+09};  
\citetalias{kvkm+03}:  \citet{kvkm+03}; 
\citetalias{mzh03}:	    \citet{mzh03};	  
\citetalias{kvk05}:	    \citet{kvk05};	  
\citetalias{kvka07}:   \citet{kvka07};  
\citetalias{vkkpm07}:  \citet{vkkpm07}; 
\citetalias{hztb04}:   \citet{hztb04};  
\citetalias{hmz+04}:   \citet{hmz+04};  
\citetalias{haberl07}: \citet{haberl07};
\citetalias{zct+01}:   \citet{zct+01};  
\citetalias{zct+05}:   \citet{zct+05};  
\citetalias{rtj+07}:   \citet{rtj+07};  
\citetalias{zmt+08}:   \citet{zmt+08};  
\citetalias{kvk09}:	    \citet{kvk09};	  
\citetalias{bhn+03}:   \citet{bhn+03};  
\citetalias{vkk07}:	    \citet{vkk07};	  
\citetalias{vkk08}:	    \citet{vkk08};	  
\citetalias{kkvk03}:   \citet*{kkvk03};  
\citetalias{vkkd+04}:  \citet{vkkd+04}; 
\citetalias{msh+05}:   \citet{msh+05};  
\citetalias{zdlmt06}:  \citet{zdlmt06}; 
\citetalias{wag+01}:   \citet{wag+01};  
\citetalias{pzs+01}:   \citet{pzs+01};  
\citetalias{dlrm03}:   \citet{dlrm03};  
\citetalias{ms02}:	    \citet{ms02};	  
\citetalias{btgg03}:   \citet{btgg03};  
\citetalias{dlcm+05}:  \citet{dlcm+05}; 
\citetalias{mgb+02}:   \citet{mgb+02};  
\citetalias{lll05}:	    \citet*{lll05};	  
\citetalias{fwa07}:	    \citet*{fwa07};	  
\citetalias{kbm+03}:   \citet{kbm+03};  
\citetalias{rn03}:	    \citet{rn03};	  
\citetalias{nrb+07}:   \citet{nrb+07};  
\citetalias{cbv+09}:   \citet{cbv+09};  
\citetalias{ghm+08}:   \citet{ghm+08};  
\citetalias{ccv+04}:   \citet{ccv+04};  
\citetalias{mpg08}:	    \citet*{mpg08};	  
\citetalias{bt97}:	    \citet{bt97};	  
\citetalias{bwt+04}:   \citet{bwt+04};  
\citetalias{zsp05}:	    \citet*{zsp05};	  
\citetalias{cgk+01}:   \citet{cgk+01};  
\citetalias{cmgc04}:   \citet{cmgc04};  
\citetalias{gkc+05}:   \citet{gkc+05};  
\citetalias{shk08}:	    \citet*{shk08};	  
\citetalias{gcm+95}:   \citet{gcm+95};  
\citetalias{gbm+99}:   \citet{gbm+99};  
\citetalias{gak+02}:   \citet{gak+02};  
\citetalias{lt08}:	    \citet{lt08};	  
\citetalias{ggg+08}:   \citet{ggg+08};  
\citetalias{ksh08}	    \citet{ksh08};	  
\citetalias{gw03}:	    \citet{gw03};	  
\citetalias{hsp+03}:   \citet{hsp+03};  
\citetalias{lwa+02}:   \citet{lwa+02};  
\citetalias{clb+02}:   \citet{clb+02};  
\citetalias{ltw08}:	    \citet*{ltw08};	  
\citetalias{km05}:	    \citet{km05};	  
\citetalias{mrg+07}:   \citet{mrg+07};  
\citetalias{gklp04}:   \citet{gklp04};  
}
\tablecomments{General pulsar data were taken from \citet{pccm02}, \citet{kfg+04}, and
 \citet{mhth05}.  Within each class the objects are ordered by
 increasing characteristic age.}
\tablenotetext{a}{Spin period $P$, dipole magnetic
 field $B_{\rm dip}=\expnt{3.2}{19}\sqrt{P {\dot P}}$, characteristic age
 $\tau_{\rm char}=P/2{\dot P}$, and spin-down energy loss rate
 $\dot E=\expnt{3.9}{46}{\dot P}/P^3$.}
\tablenotetext{b}{\textit{ROSAT} PSPC count rate (if
 known), effective temperature and radius (for spherical emission) of the
 best-fit blackbody component as measured at infinity, bolometric
 blackbody luminosity, and total luminosity in the 0.1--2.4\,keV
 band.  For sources with multiple blackbody components, we took the one
 with larger radius/smaller temperature.}
\tablenotetext{c}{Parallax distance if known.  If not, for
  the INS we use 
 distances based on X-ray absorption 
    \citep{pph+07} or 500\,pc for RX~J1308.6+2127, while for pulsars we use
 distances from  dispersion measures \citep{cl02}.}
\tablenotetext{d}{Independent age estimate.  For the INS, this is the
 kinematic age, derived from tracing the object back 
  to a probable birth site.  The range is based on the observed range
  in distance as well as the result of multiple possible birth sites.
 For pulsars this is either a kinematic age or an age estimate for the
 associated supernova remnant.}
\tablenotetext{e}{Here we use the nominal values for the spin-down from
  Table~\ref{tab:ephem}: using the upper limits does not change the conclusions.}
\tablenotetext{f}{The X-ray data are consistent with no non-thermal
 emission, but such emission cannot be ruled out.}
\tablenotetext{g}{The radius was assumed to be that of a polar cap.}
\tablenotetext{h}{The emission from this source is also consistent
 with non-thermal emission with luminosity $\sim
 \expnt{2}{29}\,\ergss$.}
\tablenotetext{i}{The X-ray source is variable, and was observed at a
  level of about $10\times$ the previous flux \citep{ggg+08,ksh08}.}
\tablenotetext{j}{The radius was fixed to that of the whole surface.}
\end{deluxetable*}
\defcitealias{kvk05}{KvK05}
\defcitealias{kvk05b}{KvK05b}
\defcitealias{vkk08}{vKK08}
\defcitealias{kvk09}{KvK09}
\defcitealias{hmz+04}{H+04}

One interesting object that may bridge the gap between the standard
pulsars and the INS is PSR~B1845$-$19: its long period and
characteristic age put it withing reach of the INS, although we do not
have an independent age or X-ray luminosity.  Compared to the INS, the
$\dot E$ for this object is slightly higher, while the characteristic
age is similar, possibly suggesting that it sits at the boundary where
magnetic field decay becomes important ($\gtrsim\!10^{13}\,$G,
according to \citealt{pmg09}).  Just how close the actual age is to
the characteristic age may prove an important test of this scenario.

\section{Conclusions}
\label{sec:conc}

We have presented initial results from coherent timing of the nearby
neutron star \rxj.  While we were not able to obtain a statistically
significant measurement of spin-down, we were able to constrain the
spin-down rate (and hence the dipolar magnetic field) to rather low values,
and with the addition of a few data points with longer time baselines
we should achieve a reliable measurement.  The limit on the magnetic
field ($<\expnt{3.7}{13}\,$G at 2\,$\sigma$) is interesting, as it is
considerably lower than the $\sim 10^{14}\,$G expected from simple
models of the X-ray spectrum.  This echoes the discrepancy seen in
\rbsb\ \citepalias{kvk09}, suggesting that we need to develop an
improved model for the X-ray spectra of the INS.  Whether this is just
from an improved treatment of the radiative transitions (e.g., auto-ionizing
transitions may play a role) or elements
beyond hydrogen we do not know, but ongoing work in obtaining better
timing constraints as well as phase-resolved spectroscopy should help to
narrow the possibilities.

The X-ray spectrum of \rxj\ shows clear signs of a broad absorption
feature at 0.3--0.6\,keV, and our data show that it
cannot be simply modeled by one or even two Gaussians.  The pulsations
are also relatively strong in the same spectral region, giving some hope
that phase-resolved spectroscopy will allow us to illuminate the
possible decompositions. 

Comparing the INS with other relevant  sub-populations of neutron
stars, we are led to the conclusion that magnetic field decay has
operated over the $\sim\!0.5\,$Myr lifetimes of the INS.  This can be
seen from a qualitative comparison of the observed X-ray luminosities
of the INS with X-ray luminosities of other sources and with
predictions from standard cooling curves.  From both, one infers that
the characteristic ages are systematically long, while the kinematic
ages are consistent with cooling models and the ages of other neutron
stars with comparable luminosities.  This conclusion is supported by
the detailed modeling of \citet{pmg09}, whose expectations for neutron
stars born with magnetic fields of $\sim2\times\!10^{14}$\,G evolved to
$\sim\!0.5\,$Myr greatly resemble the INS.  Among the implications of
this model are that some neutron stars with magnetic field decay
should be hotter at the equator than at the poles, in contrast with
most assumed models of neutron star surface temperature distributions.
Once again, phase-resolved spectroscopy would seem to be one of the best
ways to try to discriminate between the possibilities.  Improved
models for the surface temperature, which constrain the viewing
geometry, can then also be combined with proper motion measurements to try
to understand some aspects of the complicated relation between
magnetic fields, rotation, and kick velocities in young neutron stars \citep{lai01b}.

\acknowledgements We thank A.~Spitkovsky for helpful discussions.
Based on observations obtained with XMM-Newton, an ESA science mission
with instruments and contributions directly funded by ESA Member
States and NASA. DLK was supported by NASA through Hubble Fellowship
grant \#01207.01-A awarded by the Space Telescope Science Institute,
which is operated by the Association of Universities for Research in
Astronomy, Inc., for NASA, under contract NAS 5-26555.  This research
was supported in part by the National Science Foundation under Grant
No.~PHY05-51164. MHvK acknowledges funding from NSERC.
Apart from the
XMMSAS data reduction pipelines provided by \textit{XMM-Newton}, this research has made use of software provided by the
Chandra X-ray Center (CXC) in the application packages CIAO and
Sherpa.

{\it Facilities:} \facility{XMM (EPIC)}

%\bibliography{ins,inspop}

%% --------------------------------------------------------------------
%% Mon Sep 28 09:57:21 2009
%%   This file was generated automagically from the files
%%   ms.bbl and ms.tex using
%%     /Users/dlk//perl/nat2jour.pl
%%   This file should accompany ms-aas.tex.
%% --------------------------------------------------------------------

\end{document}